\documentclass[%
 reprint,
 amsmath,amssymb,
 aps,
]{revtex4-2}
\usepackage{graphicx}
\usepackage{dcolumn}
\usepackage{bm}
\usepackage{color}
\usepackage{xcolor}
\usepackage{appendix}
\usepackage{natbib}
\DeclareGraphicsExtensions{.pdf,.eps,.png,.jpg,.mps}
\begin{document}

\preprint{APS/123-QED}

\title{Self-suppressed quantum diffusion and fundamental noise limit of soliton microcombs}

\author{Xing Jin$^1$, Zhe Lv$^1$, Qihuang Gong$^{1,2,3}$, and Qi-Fan Yang$^{1,2\dagger}$}
\affiliation{$^1$State Key Laboratory for Artificial Microstructure and Mesoscopic Physics and Frontiers Science Center for Nano-optoelectronics, School of Physics, Peking University, Beijing 100871, China\\
$^2$Collaborative Innovation Center of Extreme Optics, Shanxi University, 030006, Taiyuan, China\\
$^3$Peking University Yangtze Delta Institute of Optoelectronics, Nantong, Jiangsu, 226010, China\\
}%

\begin{abstract}
Quantum diffusion of soliton microcombs has long been recognized as their fundamental noise limit. Here we surpass such limit by utilizing dispersive wave dynamics in multimode microresonators. Through the recoil force provided by these dispersive waves, the quantum diffusion can be suppressed to a much lower level that forms the ultimate fundamental noise limit of soliton microcombs. Our findings enable coherence engineering of soliton microcombs in the quantum-limited regime, providing critical guidelines for using soliton microcombs to synthesize ultralow-noise microwave and optical signals.
\end{abstract}
\maketitle

Soliton microcombs are particle-like wavepackets that have recently been generated in continuously-pumped optical microresonators \cite{Kippenberg2018}. Their formation requires a delicate balance between dispersion and nonlinearity, as well as gain and loss. The emission spectra of solitons constitute equally-spaced spectral lines, making them an ideal platform for miniature optical frequency combs \cite{chang2022integrated}. So far, soliton microcombs have demonstrated promising applications in various fields, including precision timing \cite{newman2019architecture}, microwave synthesis \cite{liang2015high,yi2017single,lucas2020ultralow,yao2022soliton}, spectroscopy \cite{suh2016microresonator,dutt2018chip,yang2019vernier}, ranging \cite{suh2018soliton,trocha2018ultrafast,riemensberger2020massively}, and parallel computing \cite{feldmann2021parallel}.

The classical perspective on the stable cyclic motions of microresonator solitons is challenged by quantum mechanics. They interact with the vacuum electromagnetic field through various loss channels of the microresonator, resulting in stochastic motion as predicted by theory \cite{matsko2013timing} and validated in experimental studies \cite{jeong2020ultralow,bao2021quantum,lao2023quantum}. The phenomenon referred to as ``quantum diffusion'' or ``quantum timing jitter'' establishes the fundamental noise limit when utilizing soliton microcombs for microwave synthesis \cite{jeong2020ultralow}. Notably, the levels of quantum diffusion of soliton microcombs are higher than those typically observed in table-top mode-locked lasers \cite{kim2016ultralow}, primarily due to the reduced modal volumes of microresonators \cite{erkintalo2021got}. Such difference in performance becomes more evident as several techniques, including pulse shaping \cite{song2011impact}, engineered dispersion \cite{kim2013sub}, and intracavity filtering \cite{shin2015characterization}, have been implemented in mode-locked lasers to minimize quantum diffusion.

Previous studies mostly focus on soliton generated in a single set of longitudinal modes. In reality, soliton-generating microresonators often support a few transverse modes. When mode coupling occurs, solitons emit dispersive waves (DWs) into other mode families, which, in turn, exert a recoil force on the solitons \cite{yi2017single,lucas2020ultralow,yang2021dispersive}. The recoil force affects the circulating velocity of the solitons within the microresonator, providing a mechanism for mitigating technical noise induced by the pump laser in repetition frequencies by counteracting Raman-self frequency shifts \cite{karpov2016raman,yi2016theory}. However, the impact of the recoil force on the quantum diffusive motions of solitons remains unclear.

\begin{figure*}
  \centering
  \includegraphics[width=\linewidth]{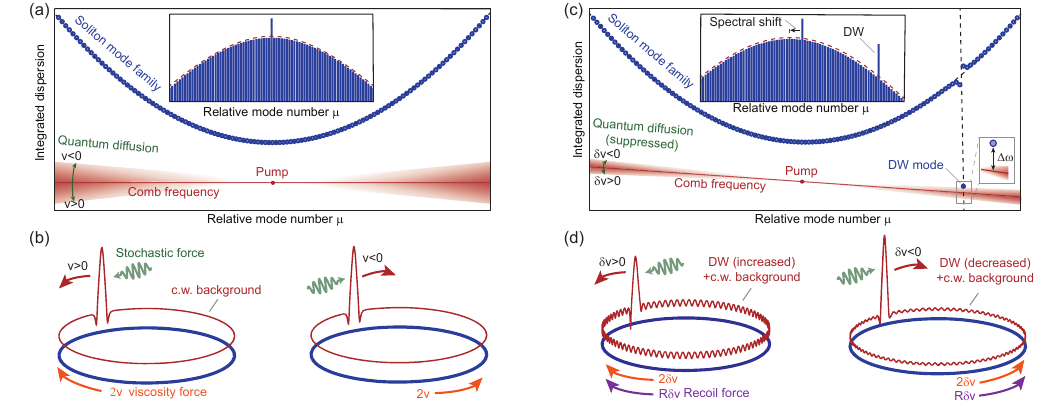}
  \caption{(a) Integrated dispersion map showing the soliton mode family (blue dots) and the comb (red line). The integrated dispersion is defined as $\omega-\omega_o-\mu D_1$, with $\omega$ the frequency of the mode (comb) and $\mu$ the order of the mode (comb) relative to the pump. The inset shows a typical optical spectrum of the soliton. (b) Quantum diffusion of solitons in the time domain. (c) Integrated dispersion map in the presence of DWs. The inset shows spectral recoil induced by the DW. (d) Quantum diffusion of solitons with DWs.}
  \label{Fig1}
\end{figure*}

In this work, we showcase the effective control of quantum diffusion of soliton microcombs using DWs. The theoretical framework starts from the dimensionless stochastic Lugiato-Lefever equation (LLE) \cite{lugiato1987spatial,coen2013modeling,matsko2013timing,herr2014temporal}:
\begin{equation}
    \frac{\partial A}{\partial \tau}=i d_2 \frac{\partial^2 A}{\partial\phi^2}+i|A|^2A-A-i\zeta A+F+\epsilon(\phi,\tau). 
    \label{Eq1}
\end{equation}
The slowly-varying envelope $A(\tau,\phi)$ at time $\tau$ and angular position $\phi$ is studied in the frame that rotates around the microresonator at the rate equal to the free spectral range ($FSR$). Note that $\tau$ is related to the lab time $t$ by $\tau=\kappa t/2$, with $\kappa$ the damping rate of the soliton mode family. The group velocity dispersion ($d_2$) and pump-resonator detuning ($\zeta$) are also normalized to $\kappa/2$. Note that the formation of solitons would often require anomalous dispersion ($d_2>0$) and a red-detuned pump ($\zeta>0$). In addition to the pump ($F$), quantum fluctuations are introduced as Langevin force $\epsilon$, which satisfies
\begin{equation}
    \langle \epsilon(\phi,\tau)\epsilon^\ast(\phi',\tau')\rangle=\frac{2\hbar\omega_o^2 n_2 D_1}{\kappa n_o A_\mathrm{eff}}\delta(\phi-\phi')\delta(\tau-\tau'),
    \label{Eq2}
\end{equation}
with $\omega_o$ the frequency of the pump mode, $n_o$($n_2$) the linear (nonlinear) refractive index, $D_1$ the $FSR$, and $A_\mathrm{eff}$ the effective mode area. The frequency-domain dynamics of the quantum diffusion is presented in Fig. \ref{Fig1}(a). The spectral components of the soliton microcombs are equally arranged in a straight line crossing the pump in the integrated dispersion map. The stochastic force randomly rotates the line about the fixed pump, changing its slope which is directly related to the repetition frequency of the soliton microcomb.

Using the single-soliton ansatz \cite{matsko2013timing, herr2014temporal}, the diffusion of the soliton in the time domain can be described by a set of equations \cite{matsko2013timing}:
\begin{equation}
\frac{\mathrm{d}x}{\mathrm{d}\tau} = v + F_x,
\label{eq:x}
\end{equation}
\begin{equation}
\frac{\mathrm{d}v}{\mathrm{d}\tau} = -2v + F_v.
\label{eq:v}
\end{equation}
In the equations, $x$ and $v$ represent the location and velocity of the soliton center in the rotation frame, respectively. The two stochastic forces, $F_x$ and $F_v$, are related to the properties of the soliton \cite{matsko2013timing}. Equation \ref{eq:v} suggests an analogy to the Brownian motion of particles suspended in a liquid. The viscosity of the liquid is provided by the microresonator damping, as illustrated in Fig \ref{Fig1}(b). Since the velocity of the soliton is associated with the shifted spectral envelope, the timing jitter attributed to $F_v$ has a mechanism similar to the Gordon-Haus jitter in optical amplifiers \cite{gordon1986random} and mode-locked lasers \cite{haus1993noise}. The Gordon-Haus jitter of the soliton is given by
\begin{equation}
    S_\mathrm{GH}(f)=\frac{\hbar \omega_o^2 n_2 \sqrt{d_2\zeta}}{6 n_o D_1 A_\mathrm{eff}(2\pi f)^2[1+(2\pi f/\kappa)^2]}.
    \label{eq:GH}
\end{equation}
Meanwhile, $F_x$ induces an instantaneous change in the location, which is known as the direct jitter. It is important to note that the dominant mechanism of quantum diffusion at timescales longer than the microresonator lifetime ($\tau\gg1$) is the Gordon-Haus jitter, which is $4\zeta^2/\pi^2$ (typically $10$-$10^2$) times greater than the direct jitter \cite{matsko2013timing}.

We then consider a secondary mode family that is linearly coupled to the soliton mode family. For simplicity, we set a significant difference in the $FSR$ between the two mode families, such that the mode interaction only occurs between a pair of modes. If the strength of the interaction is sufficient to induce avoided-mode-crossings, the frequencies of the modes will deviate from the parabolic-shaped dispersion curve. As depicted in Fig. \ref{Fig1}(c), when the hybrid mode is shifted toward the comb frequencies, the power of the comb is locally amplified, leading to the formation of a single-mode DW \cite{yi2017single}. We name this hybrid mode as the DW mode. The DW causes a spectral recoil that couples to the repetition frequency of the soliton via dispersion, resulting in a tilt of the comb frequencies in the integrated dispersion map. When the quantum diffusion causes the comb frequencies to be drawn closer to the DW mode, the increased DW enforces a larger spectral recoil that tends to push the comb frequency away, and vice versa. It is anticipated that the presence of DWs could suppress quantum diffusion compared to situations without them.

\begin{figure}
  \centering
  \includegraphics[width=\linewidth]{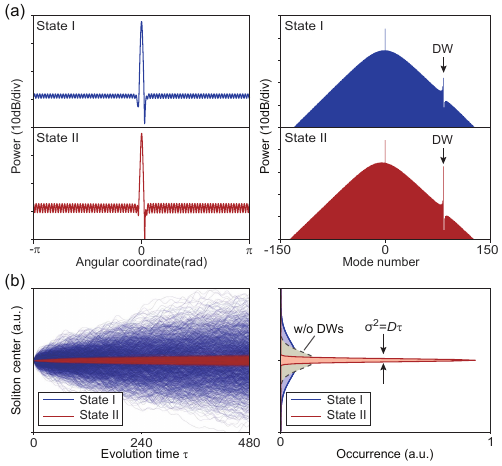}
  \caption{(a) Simulated temporal profile (left panel) and optical spectra (right panel) of different soliton states. Parameters used in generating the solitons are: $f=10.00$, $d_{2}=0.0155$, $\zeta=28.80$, and $G=65.96$. (b) Left panel: 1500 simulated traces showing the evolution of the soliton center. Right panel: distribution of the soliton center at the end of evolution. Calculated distribution w/o DW is also indicated as the gray dashed line.}
  \label{Fig2}
\end{figure}

The theory that describes the dynamics of solitons in the aforementioned situation is based on a set of coupled LLE \cite{yi2017single}, which are given by:
\begin{equation}
\frac{\partial A}{\partial \tau}=i d_2 \frac{\partial^2 A}{\partial\phi^2}+i|A|^2A-A-i\zeta A+F+\epsilon(\phi,\tau)+iGb_m e^{i m \phi},
\label{eq:CLLE_A}
\end{equation}
\begin{equation}
\frac{\mathrm{d}b_{m}}{\mathrm{d}\tau}=-\left[\eta_b+i(\omega_{b}-\omega_p - m \omega_{\mathrm{r}})\right]b_{m}+iGa_{m}+\epsilon_b(\tau).
\label{eq:CLLE_B}
\end{equation}
Here, the secondary mode ($b_m$) is coupled to the $m_\mathrm{th}$ soliton mode with coupling strength $G$. The secondary mode is characterized by its damping rate $\eta_b$ ($\eta_b=\kappa_b/\kappa$) and frequency $\omega_{b}$. The normalized pump and repetition frequencies of the soliton are denoted as $\omega_\mathrm{p}$ and $\omega_\mathrm{r}$, respectively. Equation \ref{eq:CLLE_B} represents a typical input-output relation of a linear microresonator, where the pump is provided by the $m_\mathrm{th}$ line of the soliton microcomb ($a_m=\frac{1}{2\pi}\int^{\pi}_{-\pi} Ae^{-im\phi}\mathrm{d}\phi$). Furthermore, numerical simulations presented in Appendix D confirm that the quantum stochastic force directly applied to $b_m$ ($\epsilon_b$) is negligible. Therefore, by solving the coupled LLE using the moment analysis method (see Appendix C), we obtain a modified equation of motion that describes the variation of the soliton velocity ($\delta v$) near equilibrium:
\begin{equation}
\frac{\partial \delta v}{\partial T}=- 2 \delta v-R \delta v+F_v.
\label{eq:dv}
\end{equation}
This equation introduces a recoil force due to the presence of the DW, which has the same form as the viscosity force, as illustrated in Fig. \ref{Fig1}(d). The additional viscosity term $R$ is given by:
\begin{equation}
R=\frac{4md_2\eta_b}{E}\frac{\partial |b_m|^2}{\partial \delta v}= \frac{-8md_2 \mu_c \Delta \omega}{\Delta \omega^2+\eta_b^2 }.
\label{eq:kr}
\end{equation}
where $E$ is the soliton pulse energy, $\mu_c$ is the central mode number of the soliton envelope, and $\Delta \omega$ is the detuning between the DW mode and comb line. Compared to the case without DWs, quantum diffusion can be effectively suppressed by a factor of 
\begin{equation}
    \Gamma=(1+R/2)^2.
\end{equation}
For negative $R$, it results in amplified quantum diffusion. It is worth noting that $R$ is proportional to $\partial |b_m|^2/\partial \delta v$. This suggests that a stronger DW often leads to a more significant change in quantum diffusion.


\begin{figure}
  \centering
  \includegraphics[width=\linewidth]{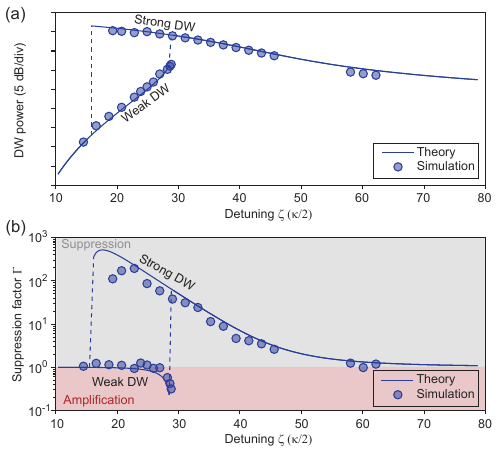}
  \caption{(a) Simulated (dots) and analytical (line) DW power versus pump-microresonator detuning. (b) Simulated (dots) and theoretical (lines) noise suppression factor as a function of detuning.}
  \label{Fig3}
\end{figure}

We numerically study the quantum diffusion of solitons in the presence of DWs using the coupled LLE. In a microresonator, we generated two soliton states with nearly identical pulse widths (see Fig. \ref{Fig2}(a)). Both states exhibited dominant DWs at the 83rd mode, although the power of the DWs differed significantly. Figure \ref{Fig2}(b) presents a comparison of the quantum diffusion of these two states by overlaying 1,500 simulated traces that depict the temporal evolution of the soliton center. The statistical analysis of the soliton center still follows a Gaussian distribution, but the full-width-at-half-maxima shows notable differences. This observation indicates that state I diffuses faster than state II. Additionally, we calculated the expected distribution of the soliton center in the absence of DWs based on Eq. \ref{Eq1} using the same parameters as states I and II \cite{matsko2013timing}. Comparing these results shows that the DWs lead to amplified (suppressed) quantum diffusion for state I (II).

\begin{figure}[hbt]
  \centering
  \includegraphics[width=\linewidth]{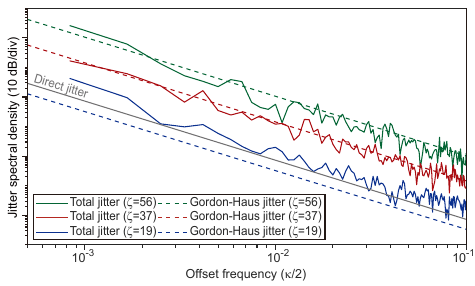}
  \caption{Spectral density of simulated total jitter (solid) and theoretical Gordon-Haus (dashed) jitter. The green, red, and blue lines correspond to solitons generated at a detuning of 55.9, 37.3, and 19.3. The level of the direct jitter for $\zeta=19.3$ is shown as the gray solid line.}
  \label{Fig4}
\end{figure}

We investigate the relationship between the DW and quantum diffusion to validate our theory. By adjusting the detuning, we access several soliton states while keeping other parameters consistent with those in Fig. \ref{Fig2}. Figure \ref{Fig3}(a) shows a bistable behavior of the DW power within the detuning range of 16.6 and 28.8 \cite{yi2017single}. This leads to the emergence of two branches of soliton states, with one branch featuring DWs with significantly higher power. Notably, for detunings between 47.7 and 53.8, the solitons become unstable due to the formation of the DW, known as the intermode breather solitons \cite{guo2017intermode}. To analyze the impact of DWs on quantum diffusion, we simulate the quantum diffusion rates for stable solitons and compare them with the case without DWs, aiming to derive the noise suppression factor $\Gamma$. The results, presented in Fig. \ref{Fig3}(b), align closely with the theoretical predictions based on Eq. \ref{eq:kr}. It should be noted that solitons on the strong-DW branch exhibit suppressed quantum diffusion, whereas those on the weak-DW branch manifest amplified quantum diffusion. This phenomenon can be elucidated by the qualitative analysis illustrated in Fig. \ref{Fig1}(c). Specifically, for the strong-DW branch, the larger spectral shifts result in a steeper comb frequency slope, which tends to position itself below the DW mode. As a result, this situation yields a positive $\partial |b_m|^2/\partial \delta v$, consequently leading to a positive $R$. Conversely, the weak-DW branch is associated with a negative $R$ caused by the comb frequency located above the DW mode. It is worth noting that a maximum reduction factor of 20.8 dB is attained when the detuning is set to 19.3.

One final question remains: is there a limit to the suppression of quantum diffusion? We note that the direct jitter is not dependent on the velocity of the soliton. Therefore, the mechanism that suppresses the Gordon-Haus jitter is not applicable to the direct jitter. The spectral density of direct jitter is given by:
\begin{equation}
    S_\mathrm{D}(f)=\frac{\pi^2 \hbar \omega_o^2 n_2 \sqrt{d_2}}{ 24n_o D_1 A_\mathrm{eff}\zeta^{3/2}(2\pi f)^2}.
    \label{eq:direct}
\end{equation}
To test this limit, we have selected several states with different suppression factors. As depicted in Fig. \ref{Fig4}, for small suppression factors, the timing jitter aligns with the Gordon-Haus jitter. However, we observe a deviation from this agreement when the Gordon-Haus jitter is suppressed by 26 dB to a level lower than the direct jitter. In such cases, the primary limitation on timing jitter arises from the direct jitter, which should be considered the ultimate noise limit for soliton microcombs.

In contrast to the Gordon-Haus jitter, it is anticipated that the direct jitter will decrease as the detuning increases. This challenges the conventional belief that broadband soliton microcombs generated with a large detuning will experience significant quantum timing jitter. Moreover, the notion that the noise performance of soliton microcombs can never match that of table-top mode-locked lasers is being questioned. Theoretical calculations predict that the level of direct jitter for soliton microcombs generated in Si$_3$N$_4$ microresonators at 1550 nm wavelength can reach $1\times 10^{-8} $ fs$^2$/Hz at 10 kHz offset frequency, given the following parameters $n_2=2.2\times 10^{-19}$m$^2$/W, $n_o=2$, $A_\mathrm{eff}=2$ $\mu$m$^2$, $d_2=0.001$, $\zeta=60$, $D_1=2\pi\times 100$ GHz. This level is on par with that of state-of-the-art mode-locked lasers \cite{kuse2016all}. However, it should be noted that achieving direct-jitter-limited performance may pose practical challenges. Equation \ref{eq:kr} suggests that a significant suppression factor may require extremely high-power DWs. Nevertheless, this could potentially induce a substantial spectral shift in the soliton, leading to breathing-type instabilities \cite{guo2017intermode} or even preventing soliton formation \cite{herr2014mode}. Some mechanisms that might help reduce the overall spectral shift include Raman effects \cite{skryabin2003soliton,yi2016theory,yi2017single} or the utilization of multiple DWs.

In conclusion, we have demonstrated the possibility of self-suppressing quantum diffusion in soliton microcombs through the emission of DW radiation. This mechanism can also be applied to microcombs produced in normal dispersion microresonators, leading to even lower levels of noise thanks to increased pulse energy \cite{lao2023quantum}. Our findings extend the coherence engineering of soliton microcombs into the quantum regime. The resulting ultralow-noise microcombs have the potential to revolutionize high-precision metrology provided their compelling advantages in size, weight, and power consumption. 

The authors thank Lu Yao and Binbin Nie at Peking University for their helpful discussions. The project is supported by National Key R\&D Plan of China (Grant No. 2021YFB2800601), Beijing Natural Science Foundation (Z210004), National Natural Science Foundation of China (92150108), and the High-performance Computing Platform of Peking University.

Xing Jin and Zhe Lv contributed equally to this work.

\appendix
\section{Quantum diffusion of solitons w/o dispersive waves}
In this section, we provide a theoretical analysis of the quantum diffusion dynamics of solitons in the absence of dispersive waves. Our theoretical framework is based on the Lugiato-Lefever Equation (LLE) in addition to an additional stochastic force term to account for quantum noise \cite{matsko2013timing}:
\begin{equation}
\begin{aligned}
    \frac{\partial \psi}{\partial t}&=i\frac{D_{2}}{2}\frac{\partial^2 \psi}{\partial \phi^2}+ig|\psi|^2 \psi -\frac{\kappa}{2}\psi-i\delta\omega \psi\\
    &+\sqrt{\frac{\kappa_\mathrm{ext}P_\mathrm{in}}{\hbar \omega_o}}+F_q(\phi,t).
\end{aligned}
\label{LLE}
\end{equation}
\begin{equation}
<F_q(\phi,t)F_q^*(\phi',t')>=\pi \kappa \delta(\phi-\phi')\delta(t-t').
\label{LLE_sto}
\end{equation}
with $\psi$ the intracavity field normalized to the photon number, $D_2$ the second-order dispersion of the microresonator, $g=\frac{\hbar\omega_{0}^2 n_{2}D_{1}}{2\pi n_{0}A_\mathrm{eff}}$ the nonlinear factor, $\kappa$ the total decay rate of the soliton mode family, $\kappa_\mathrm{ext}$ the coupling coefficient with the pump, $P_\mathrm{in}$ the pump power, and $F_q(\phi,t)$ the stochastic force originated from the vacuum fluctuations. Equation \ref{LLE} and \ref{LLE_sto} can be further normalized by setting $d_2 =D_{2}/ \kappa$, $A=\sqrt{2g/\kappa}\psi$, and $\tau=\kappa t/2$:
\begin{equation}
    \frac{\partial A}{\partial \tau}=id_{2}\frac{\partial^2 A}{\partial \phi^2}+i|A|^2 A -A-i\zeta A+F+\epsilon(\phi,\tau),
\label{normlle}
\end{equation}

\begin{equation}
    \langle \epsilon(\phi,\tau)\epsilon^\ast(\phi',\tau')\rangle=\frac{2\hbar\omega_o^2 n_2 D_1}{\kappa n_o A_\mathrm{eff}}\delta(\phi-\phi')\delta(\tau-\tau').
    \label{nLLE_sto}
\end{equation}
with $\zeta=2\delta\omega/\kappa$ the normalized detuning, and $F=\sqrt{8g\kappa_\mathrm{ext}P_\mathrm{in}/\kappa^3\hbar\omega_o}$ the normalized pump term. These equations correspond to Eq. 1 and 2 in the main text.

A straightforward description of soliton quantum diffusion can be obtained by using the moment analysis method. The moments under consideration are the soliton energy
\begin{equation}
    E=\frac{1}{2\pi}\int_{-\pi}^{\pi}|A|^2\mathrm{d}\phi,
    \label{E}
\end{equation}
the location of the soliton center
\begin{center}
\begin{equation}
\begin{split}
x=\frac{1}{2\pi E}\int_{-\pi}^{\pi}\phi|A|^2 \mathrm{d}\phi,
\end{split}
\label{phic}
\end{equation}
\end{center}
and the central mode number of the soliton envelope
\begin{center}
\begin{equation}
\begin{split}
\mu_c =\frac{\sum_{\mu}\mu|a_{\mu}|^2}{E}=\frac{1}{4\pi i E}\int_{-\pi}^{\pi}(A^{*}\frac{\partial A}{\partial \phi}-A\frac{\partial A^* }{\partial \phi})\mathrm{d}\phi.
\end{split}
\label{muc}
\end{equation}
\end{center} 
where $a_\mu=\frac{1}{2\pi}\int_{-\pi}^{\pi}A(\phi)e^{-i\mu\phi}d\phi$ is the amplitude of the $\mu$th mode. The center mode number $\mu_c$ is related to the group velocity of solitons due to the microresonator dispersion. By taking the time derivatives of Eq. \ref{phic} and \ref{muc} and combing with Eq. \ref{normlle}, we can get a set of motion equations: 

 \begin{equation}
     \frac{{\mathrm{d}}x}{\mathrm{d} \tau}=v+F_x,
     \label{eqx}
 \end{equation}

  \begin{equation}
     \frac{{\mathrm{d}}v}{\mathrm{d} \tau}=-2v+F_v.
     \label{eqv}
 \end{equation}
Here, $v=2d_2\mu_c$ presents the velocity of solitons. The stochastic forces $F_x$ and $F_v$ are defined as:

\begin{equation}
    F_{x}=\frac{1}{2\pi E}\int_{-\pi}^{\pi}\phi(A\epsilon^{*}+A^{*}\epsilon)\mathrm{d}\phi,
    \label{Fx1}
\end{equation}

\begin{equation}
    F_{v}=\frac{d_2}{\pi iE}\int_{-\pi}^{\pi}(\epsilon^{*}\frac{\partial A}{\partial \phi}-\epsilon \frac{\partial A^{*}}{\partial \phi})\mathrm{d}\phi.
    \label{Fv1}
\end{equation}
By utilizing the single soliton ansatz, given as $A(\phi)=\sqrt{2\zeta}\mathrm{sech(\phi\sqrt{\zeta/d_2})}$ \cite{matsko2013timing,herr2014temporal}, we can get:
\begin{equation}
    <F_x(\tau)F_x(\tau')>=\frac{\pi^2\hbar\omega_o^2n_2 D_1\sqrt{d_2}}{12\kappa n_oA_\mathrm{eff}\zeta^{3/2}}\delta(\tau-\tau'),
\end{equation}
\begin{equation}
    <F_v(\tau)F_v(\tau')>=\frac{4 n_2 \sqrt{d_{2}\zeta }\hbar\omega_{o}^2 D_{1}}{3\kappa^2 n_{0}A_\mathrm{eff}}\delta(\tau-\tau'),
\end{equation}
The quantum timing jitter power spectral density can be further derived as follows:
\begin{equation}
\begin{aligned}
    S_t(f)=\frac{S_x(f)}{D_1^2}&=\frac{\hbar \omega_o^2 n_2 \sqrt{d_2\zeta}}{24\pi^2 n_o D_1 A_\mathrm{eff}f^2}[\frac{1}{1+(2\pi f/\kappa)^2}\\
    &+\frac{\pi^2}{4\zeta^2}].
\end{aligned}
\label{PSD1}
\end{equation}
The first term on the right-hand side of Eq. \ref{PSD1} corresponds to the Gordon-Haus jitter power spectral density, which is identical to Eq. 5 in the main text. The subsequent term represents the direct jitter PSD, as expressed in Eq. 11 in the main text. At low offset frequency ($f\ll\kappa$), the Gordon-Haus jitter is $4\zeta^2/\pi^2$ times greater than the direct jitter and the expression can be simplified as:
\begin{equation}
    S_{GH}(f)\approx\frac{\hbar \omega_o^2 n_2 \sqrt{d_2\zeta}}{6 n_o D_1 A_\mathrm{eff}(2\pi f)^2}.
    \label{GH0}
\end{equation}

\section{Soliton dynamics with single-mode dispersive waves}
In this section, we discuss the dynamics of solitons in the presence of dispersive waves. For simplicity, we study a scenario where the dispersive wave mode family (mode family B) is linearly coupled with the soliton-support mode family (mode family A), with the dispersive waves predominantly focusing on a specific mode\cite{yi2017single}. In such a case, a set of equations can be applied to describe this model:

\begin{center}
\begin{equation}
\begin{split}\frac{\partial A(\phi,t)}{\partial \tau}=i{d_2}\frac{\partial^2 A}{\partial \phi^2}+i|A|^2 A -A-i\zeta A+F+iGb_{m}e^{im\phi},
\end{split}
\end{equation}
\end{center}
\begin{equation}
\begin{split}\frac{\mathrm{d}b_{m}}{\mathrm{d}\tau}=&-[\eta_{b}+i(\omega_{b}-\omega_{p}-m\omega_{r})]b_{m}+iGa_{m}.
\end{split}
\label{bmode}
\end{equation}
with $G$ representing the linear coupling coefficient, $\omega_p$ as the pump frequency, $\omega_r$ as the repetition rate of microcombs, and $\omega_{b}$ as well as $\eta_{b}=\kappa_B/\kappa$ signifying the frequency and dissipation of the $m_{th}$ mode within mode family B. The coupling of the $m_{th}$ mode in mode family A and B results in the formation of hybrid modes and the frequencies of the upper and lower modes can be expressed as \cite{liu2014investigation,haus1991coupled}:

\begin{equation}
\omega_{\pm}=\frac{\omega_{a}+\omega_{b}}{2}\pm \sqrt{G^2+\frac{(\omega_{a}-\omega_{b})^2}{4}}.
\end{equation}
with $\omega_{a}$ the frequency of the $m_{th}$ mode in mode family A. The amplitude of the hybrid modes is the linear combination of $a_m$ and $b_m$. When the dispersive wave mode is significantly separated from the pump, there will be:
\begin{equation}
   |\omega_{a}-\omega_{b}|\gg G,
   \label{app}
\end{equation}
A straightforward approximation of the amplitude of the lower hybrid mode can be made:
\begin{equation}
h_{m-}=\frac{G a_{m}+(\omega_{a}-\omega_{b})b_m}{\sqrt{G^2 +(\omega_{a}-\omega_{b})^2}}.
\label{hybrid}
\end{equation}
Take the time derivative of Eq. \ref{hybrid} and we can get:

\begin{equation}
    \frac{d h_{m-}}{d \tau}=-(\eta_{-}+i\Delta \omega) h_{m-}+F_m.
\end{equation}
where $\eta_{-}$ is the decay rate of the lower hybrid mode and it is approximately equal to $\eta_{b}$ under the condition of Eq. \ref{app}.  
$\Delta \omega=\omega_{-}-\omega_{p}-m\omega_r$ is the relative detuning between the lower hybrid mode and the $m_{th}$ comb line. Considering $\omega_r=d_1+2\mu_c d_2$ ($d_1$ the normalized FSR of mode family A), $\Delta \omega$ can also be presented as $\omega_{-}-\omega_p-m d_1-mv$. $F_m$ is the pump for the lower hybrid mode which can be expressed as \cite{yi2017single}:
\begin{equation}
\begin{split}
F_{m}=i\gamma(\omega_{a}-\omega_p-md_1-mv)a_{m}.
\end{split}
\label{lowbranch}
\end{equation}
where $\gamma = G/\sqrt{G^2+(\omega_{a}-\omega_{b})^2}\ll 1$. From Eq. \ref{lowbranch}, the stable solution for the dispersive amplitude can be obtained as follows:
\begin{equation}
\begin{split}
h_{m-}=\frac{F_m}{i\Delta\omega+\eta_{b}}.
\end{split}
\label{DWpower}
\end{equation}
Considering Eq. \ref{app}, we can get:
\begin{equation}
    b_m\approx \frac{h_{m-}}{\sqrt{1-\gamma^2}}=\frac{F_m}{\sqrt{1-\gamma^2}(\eta_{b}+i\Delta\omega)}.
    \label{bmpower}
\end{equation}
By using the moment analysis method, we can get soliton motion dynamics in the presence of single-mode dispersive waves:
\begin{equation}
\begin{split}
\frac{\mathrm{d} v}{\mathrm{d} \tau}=
-2 v +F_{DW},
\end{split}
\label{Fmuc}
\end{equation}
where $F_{DW}$ represents the recoil force exerted by dispersive waves on solitons, and the detailed expression is as follows:
\begin{equation}
\begin{aligned}
F_{DW}=&-\frac{2d_2}{2\pi iE}\int_{-\pi}^{\pi}(iGb_{m}(-im)a_{m}^{*}+iGb_{m}^{*}(im)a_{m})\mathrm{d}\phi
\\=&-\frac{4md_2\eta_{b}|b_{m}|^2}{E}. 
\end{aligned}
\end{equation}
Equation \ref{bmode} is used in the above calculations. Solving Eq. \ref{Fmuc}, we can observe that the recoil force imparts a stable velocity to the soliton, which is denoted as $v_o$:
\begin{equation}
    v_o=-\frac{2md_2\eta_{b}|b_{m}|^2}{E}.
    \label{vo}
\end{equation}
The non-zero velocity also corresponds to the central mode
number of the soliton envelope deviating from the pump, which is denoted as $\mu_c=v_o/2d_2=-m\eta_{b}|b_m|^2/E$. Combining Eq. \ref{bmpower} and \ref{vo}, we can get the equation of dispersive wave power:
\begin{equation}
    |b_{m}|^2=\frac{|F_m|^2}{(1-\gamma^2)[\eta_{b}^2+(\Delta\omega_o+2m^2d_2\eta_{b}|b_{m}|^2/E)^2]}.
    \label{PDW1}
\end{equation}
Here, $\Delta\omega_o=\omega_{-}-\omega_p-md_1$. Equation \ref{PDW1} represents a cubic equation concerning the dispersive wave power. In specific parameter configurations, multiple solutions can exist, and these solutions correspond to bistable states, as depicted in Fig. 3 in the main text.

\section{Quantum diffusion of solitons with single-mode dispersive waves}
In this section, we study the quantum diffusion dynamics of solitons with single-mode dispersive waves. The theoretical model is based on a set of the Lugiato-Lefever equations (LLEs) with two stochastic-force terms:
\begin{center}
\begin{equation}
\begin{split}\frac{\partial A(\phi,\tau)}{\partial \tau}=&i{d_2}\frac{\partial^2 A}{\partial \phi^2}+i|A|^2 A -A-i\zeta A+F
\\&+iGb_{m}e^{im\phi}+\epsilon(\phi,\tau),
\end{split}
\end{equation}
\end{center}
\begin{equation}
\begin{split}\frac{\mathrm{d}b_{m}}{\mathrm{d}\tau}=&-[\eta_{b}+i(\omega_{b}-m\omega_{r}-\omega_{p})]b_{m}+iGa_{m}+\epsilon_{b}(\tau).
\end{split}
\label{bmode1}
\end{equation}
where $\epsilon_{b}$ satisfies:
\begin{equation}
<\epsilon_{b}(\tau)\epsilon_{b}^{*}(\tau^{'})>=\frac{\hbar\omega_{o}^{2}n_{2}D_{1}}{\pi\kappa n_{o} A_\mathrm{eff}}\delta(\tau-\tau^{'}).
\end{equation}
The influence of $\epsilon_b$ on the quantum diffusion process of solitons is found to be negligible, as demonstrated in the subsequent simulations. Therefore, we will not take into account the impact of $\epsilon_b$ in the following discussions. Utilizing the same moment method as in Section I, we can derive the following dynamic equations:
\begin{center}
\begin{equation}
\begin{split}
\frac{{\mathrm{d}}x}{\mathrm{d} \tau}=v+F_{x},
\label{dwsx}
\end{split}
\end{equation}
\end{center}
\begin{equation}
\begin{split}
\frac{\mathrm{d} v}{\mathrm{d}\tau}=-2 v -\frac{4md_{2}|b_m|^{2} \eta_{b}}{E}+F_{v}.
\end{split}
\label{dws}
\end{equation}
where the expressions for $F_x$ and $F_v$ are given in Eq. \ref{Fx1} and \ref{Fv1}. The expression for $|b_m|^2$ at a steady state is as Eq. \ref{PDW1}. It is worth noting the value of $|b_m|^2$ is related to the soliton velocity $v$. Solving Eq. \ref{dws} directly can be challenging and cumbersome. However, since our primary focus is on quantum diffusion over timescales longer than the microresonator's lifetime (at offset frequencies much smaller than $\kappa$, where Gordon-Haus jitter dominants), we can consider $|b_m|^2$ as a quasi-steady state and take the following linear approximations:

\begin{equation}
\begin{split}
 v=v_{o}+\delta v,
\end{split}
\label{dv}
\end{equation}

\begin{equation}
\begin{split}
 x=v_{o}\tau+\delta x,
\end{split}
\label{dx}
\end{equation}

\begin{equation}
\begin{split}
 |b_m|^2=|b_{mo}|^2+\frac{\partial|b_m|^2}{\partial \delta v}\delta v.
\end{split}
\label{db}
\end{equation}
By substituting Eq. \ref{dv}, Eq. \ref{dx}, and Eq. \ref{db} into Eq. \ref{dwsx} and Eq. \ref{dws}, we obtain:
\begin{equation}
    \frac{d\delta x}{d\tau}=\delta v+F_x,
    \label{ddx}
\end{equation}
\begin{equation}
    \frac{d\delta v}{d\tau}=-2\delta v-R\delta v+F_v.
    \label{ddv}
\end{equation}
where $R=\frac{4md_{2} \eta_{b}}{E}\frac{\partial |b_m|^{2}}{\partial \delta v}$ represents the dispersive waves-induced additional viscosity. To get the expression of additional viscosity $R$, we can rewrite Eq. \ref{PDW1} as:
\begin{equation}
    |b_m|^{2}=\frac{|F_{m}|^2}{(1-\gamma^2)[\eta_{b}^{2}+(\Delta\omega-m\delta v)^2]}.
\end{equation}
Then we can get:
\begin{center}
\begin{equation}
\begin{split}
R&=\frac{4md_{2}\eta_{b}}{E}\frac{\partial |b_m|^{2}}{\partial \delta v}
\\&=-\frac{8md_{2}\mu_{c}\Delta\omega}{\eta_b^{2}+\Delta\omega^{2}}-4d_2\mu_{c}\frac{\partial |F_{m}|^2}{|F_{m}|^2\partial \delta v}
\\&\approx -\frac{8md_{2}\mu_{c}\Delta\omega}{\eta_b^{2}+\Delta\omega^{2}}.
\end{split}
\label{viscosity}
\end{equation}
\end{center}
The relationship $\mu_c=-m\eta_b|b_{mo}|^2/E$ is used in the above calculation. Combining equation \ref{ddx} and \ref{ddv}, the quantum timing jitter spectral density of solitons with a single-mode dispersive wave can be calculated as:
\begin{equation}
\begin{aligned}
       S_t(f)=\frac{S_x(f)}{D_1^2}&=\frac{\hbar \omega_o^2 n_2 \sqrt{d_2\zeta}}{24\pi^2 n_o D_1 A_\mathrm{eff}f^2}[\frac{1}{(1+\frac{R}{2})^2+(2\pi f/\kappa)^2}\\
    &+\frac{\pi^2}{4\zeta^2}]. 
\end{aligned}
\label{PSDwd}
\end{equation}
It can be seen from Eq. \ref{PSDwd} that the presence of single-mode dispersive waves can suppress the Gordon-Haus jitter, while the direct jitter can not be suppressed and hence becomes the ultimate noise limit. At low offset frequency $f\ll\kappa$:
\begin{equation}
\begin{aligned}
       S_{GH}(f)\approx\frac{\hbar \omega_o^2 n_2 \sqrt{d_2\zeta}}{24\pi^2 n_o D_1 A_\mathrm{eff}f^2(1+R/2)^2}.
\end{aligned}
\label{GH1}
\end{equation} 
When comparing Eq. \ref{GH0} and Eq. \ref{GH1}, the suppression factor $\Gamma$ can be defined as:
\begin{equation}
    \Gamma=\frac{1}{(1+R/2)^2}.
\end{equation}

\section{Impact of noise terms}
\begin{figure}[h]
  \centering
  \includegraphics[width=\linewidth]{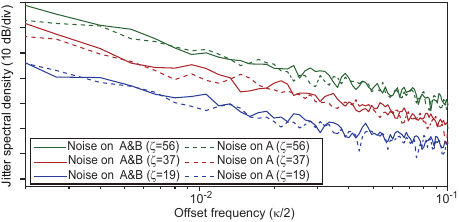}
  \caption{Simulated jitter spectral density of three soliton states with quantum noise on both mode families A and B (solid lines) as well as solely on mode families A (dashed lines). The green, red, and blue lines correspond to solitons generated at the detuning of 56, 37, and 19.}
  \label{SIFig1}
\end{figure}

The above theoretical analysis neglects the quantum noise directly applied to $b_m$. Here, we perform numerical simulations to validate this approximation. Quantum noise was applied to both mode families A and B, as well as exclusively to mode family A in the simulations of three different soliton states with varying DWs power. The results are presented in Fig \ref{SIFig1}. These three soliton states exhibit different suppression factors. Remarkably, the simulated jitter spectral density traces, with and without quantum noise on mode family B, exhibit a high degree of similarity in all three states. This consistency serves to validate our approximation. 

\bibliography{ref}
\end{document}